# Title: Collisional and Rotational Disruption of Asteroids

**Authors:** Kevin J. Walsh (Observatoire de la Côte d'Azur), Patrick Michel (Observatoire de la Côte d'Azur) and Derek C. Richardson (University of Maryland)

**Correspondence:**
Kevin J. Walsh
Observatoire de la Côte d'Azur
UMR6202 Cassiopée/CNRS
Nice, France 06304
Tel: (33) 6 9200 1944
Email: kwalsh@oca.eu

**Abstract**:

Asteroids are leftover pieces from the era of planet formation that help us understand conditions in the early Solar System.  Unlike larger planetary bodies that were subject to global thermal modification during and subsequent to their formation, these small bodies have kept at least some unmodified primordial material from the solar nebula. However, the structural properties of asteroids have been modified considerably since their formation.  Thus, we can find among them a great variety of physical configurations and dynamical histories. In fact, with only a few possible exceptions, all asteroids have been modified or completely disrupted many times during the age of the Solar System.  This picture is supported by data from space mission encounters with asteroids that show much diversity of shape, bulk density, surface morphology, and other features.  Moreover, the gravitational attraction of these bodies is so small that some physical processes occur in a manner far removed from our common experience on Earth.  Thus, each visit to a small body has generated as many questions as it has answered.  In this review we discuss the current state of research into asteroid disruption processes, focusing on collisional and rotational mechanisms.  We find that recent advances in modeling catastrophic disruption by collisions have provided important insights into asteroid internal structures and a deeper understanding of asteroid families.  Rotational disruption, by tidal encounters or thermal effects, is responsible for altering many smaller asteroids, and is at the origin of many binary asteroids and oddly shaped bodies.

## Introduction:

Asteroids are the remnants of planet formation, making them a subject of great interest.  Studying their current properties and orbits provides a snapshot of a population of bodies with vast physical variety that has been evolving since the planets formed, around 4.5 Gyr ago.  In order to better understand the early stage of planet formation, it is instructive to determine the mechanisms that continue to evolve the asteroids today.  Thanks to modern innovations in asteroid observations, researchers have information on millions of bodies, ranging in diameter from meters to 1000's of km and spanning nearly an order of magnitude in density, rotating in minutes or months, showing surfaces as dark as charcoal or nearly as bright as snow, with solid metallic cores or massive void spaces, and that have left their mark on terrestrial worlds by way of impact craters.

The scope of this review is limited to the collisional and rotational disruption of asteroids in the near-Earth, Main Belt and Jupiter Trojan populations. We focus on the most recent research and try to put into context how new results have changed our understanding of the asteroid population.

**Asteroids**

The most general classification of asteroids is based on dynamical considerations, or roughly where in the Solar System they spend most of their time. The closest group to the Sun is composed of so-called near-Earth asteroids that evolve on orbits that can cross or come close to Earth's (perihelion distances $q \leq 1.3$ and aphelion distances $Q \geq 0.983$ [1]). Near-Earth asteroids (NEAs) range in physical size from meters to around 10 km (about 1000 NEAs have diameters larger than 1 km), have very short dynamical lifetimes of around 10 Myr, and are expected to only rarely collide with each other.[2,3] Their lifetime is limited due to encounters with planets, or injection into resonances that increase their eccentricity until encounters with Jupiter eject them from the Solar System, or they collide with the Sun. Most major asteroid taxonomic classifications are represented among the NEAs, though they are not strictly a reflection of the Main Belt between Mars and Jupiter; these bodies mostly inhabit the inner part of the Main Belt, making migration to the near-Earth population slightly easier. Because of their limited lifetimes, NEAs cannot have formed in their present locations, but rather must have migrated into their relatively chaotic and short-lived orbits from a few major sources that have been identified in recent years (e.g. [1]). The first one is the large reservoir of asteroids between the orbits of Mars and Jupiter: the Main Belt.

Main Belt asteroids (MBAs) are physically very diverse, with sizes ranging up to 1000 km, densities spanning between ~0.5 g cm$^{-3}$ to ~5 g cm$^{-3}$, and compositions of loosely bound porous material or solid differentiated structures with metallic cores and basaltic crusts. Most are dynamically stable, with lifetimes comparable to the age of the Solar System. However, their lifetimes against collisional disruption are size dependent, and for most (probably those with diameter smaller than 50 km) are shorter than the age of the Solar System.[4] Thus it is expected, and largely confirmed by the growing number of known collisional asteroid families, that the Main Belt is collisionally evolved. The thermal Yarkovsky effect, which causes a drift in semi-major axis on timescales that depend on the size of the body (decreasing with smaller size) and its distance to the Sun (decreasing with distance), increases the mobility of smaller bodies (below ~10 km), allowing them to be steadily injected into the resonances that feed the NEA population, keeping the number of bodies roughly in a steady-state.

The Jupiter Trojans are a numerous group of asteroids that share Jupiter's orbit. They have similar orbital semi-major axes as Jupiter, and reside in clouds 60 degrees ahead of and behind Jupiter, near the $L_4$ and $L_5$ Lagrange points. This extremely large population is more physically homogeneous than either of the two inner populations (NEAs and MBAs), and is dominated by primitive materials. Recent numerical simulations suggest that bodies of this population come from a disk of planetesimals that was evolving beyond the orbits of the newly formed giant planets: the Trojans were captured during a highly unstable period when Jupiter and Saturn came into mutual resonance, leading to the emplacement of Uranus and Neptune into their current orbits and to a massive injection of bodies into the inner Solar System (the Late Heavy Bombardment).[5] The Trojan population is limited in the volume of space it occupies near the ecliptic plane, but does have a high mean inclination that helps to keep its collisional activity similar to that of the Main Belt.[6] Recent numerical models suggest that in the past the Trojans were more collisionally active than they are today and successfully predicted some of the observed collisional families within this population.[7,8]

Kuiper Belt objects (KBOs), which will only be mentioned in passing, inhabit the space around and beyond the orbit of Neptune. Although suspected to exist for decades,[9,10] this population was only recently discovered observationally (the discovery of the first object, 1992 QB1, was by [11]) and many of its basic physical and dynamical properties are still under investigation. It is believed most KBOs are cometary in nature.

One of the strongest constraints on the collisional history of the MBAs and NEAs is their size distribution. The details reveal the current state of the asteroid populations, but the basic features of the size distribution, and deviations from the mean slope, tell a story about their long-term evolution. If all bodies of all sizes disrupted in exactly the same manner, the size distribution would evolve to a stationary solution and a simple power law.[12] However, the real size distribution of both MBAs and NEAs departs from such a simple power law (Fig. 1) and some features are interpreted as being due to dependence of collisional outcomes on body size and composition. Such a dependency is suggested from collisional experiments at small scales and numerical simulations at larger scales (see Sec. 1).

Recent studies indicate that the size distribution of large MBAs (> 100 km in diameter) and some features of the full distribution (such as bumps) may be primordial and may not have been altered much after the Late Heavy Bombardment.[4] Nonetheless, there are still uncertainties in this distribution and some aspects that are not clearly understood, especially at small sizes (less than a few km), below the observational completeness limit. Models of the size distribution are quite sensitive to changes in disruption models and remain one of the strongest constraints for any new work on disruption outcomes. The upcoming deluge of asteroid discoveries and characterizations from very deep all-sky surveys (LSST and Pan-Starrs), will extend the distribution to very small sizes and improve the completeness at larger sizes.

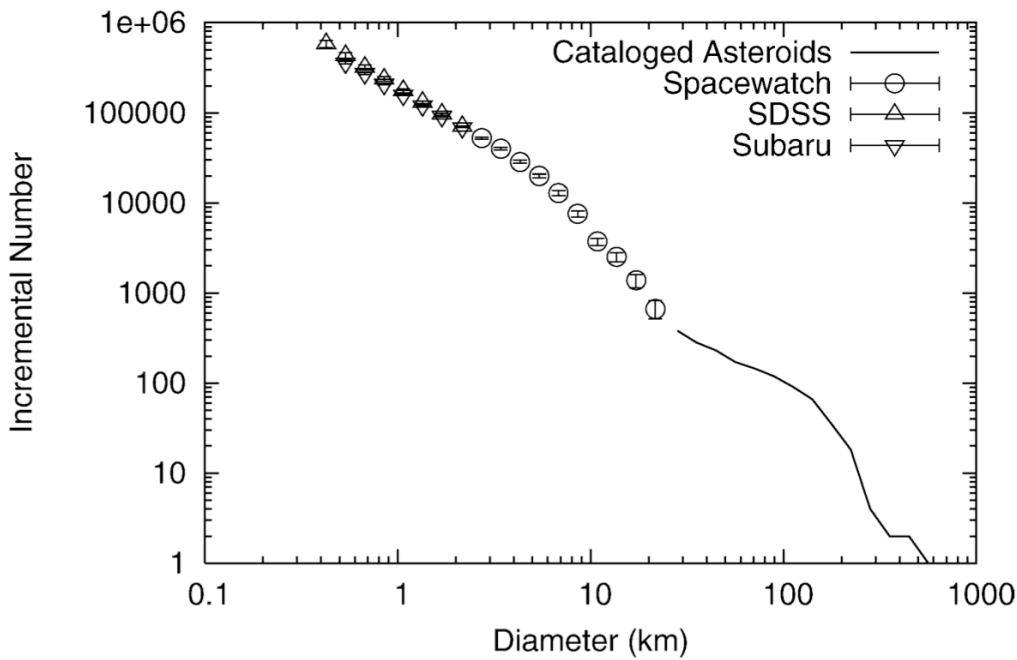

**Figure 1**: The size distribution of Main Belt asteroids, with the cumulative number larger than a given diameter plotted against diameter (km). Many different determinations of this relationship are shown, each drawing data from different sources. The larger trend emerges with two bumps, around ~5 km and ~100 km.

# 1. Collisional Disruption

Once the planets were largely formed, relative speeds between small bodies increased as a result of planetary perturbations. Consequently, our Solar System entered a new regime of high impact energy, in which it continues to evolve. In July 1994, observations of the impact of comet Shoemaker Levy 9 on Jupiter showed that collisions continue to occur. Furthermore, meteorites collected on Earth are the remnants of very recent collisions in the main asteroid belt, and impacts of larger bodies with Earth are a real threat to our biosphere. The collisional process is therefore not a second-order problem in the understanding of the past, present and future history of our Solar System; it is actually at the heart of its formation and evolution.

The scales of the phenomena that are involved in planetary and small-body impacts are by far much larger than those reached in laboratory experiments. Extrapolations by 15 orders of magnitude in mass are typically necessary to achieve ranges that are relevant to asteroids and planetesimals. However, laboratory experiments are crucial to validate the numerical simulations of fragmentation at small scale before they are applied at larger scales. These experiments are done using tools capable of launching projectiles at speeds corresponding to impacts in the asteroid belt (about 5 km/s),[3] commonly a two-stage light gas gun. A target is placed in a chamber and cameras are oriented at two angles for filming the event. The fragment mass distribution is determined by recovering and weighing the fragments, and the velocity distribution is measured from the movies. Since only the largest fragments can be easily recognized in the movies, data on ejection velocities are still sparse. The validation of numerical codes relies on their ability to reproduce the measurable properties using the same initial conditions and material parameters, though the latter can be difficult to characterize.

Researchers distinguish between the *strength* and *gravity* regimes when discussing impact scenarios. The strength regime applies to impacts between objects whose mechanical binding energy exceeds their gravitational binding energy, so that the effect of gravity can largely be neglected. Thus, in the strength regime, the main parameter that influences the outcome of the collision is the material strength of the body. Laboratory measurements indicate that larger bodies are weaker because they have a higher probability to contain larger incipient flaws that decrease their strength.[14] The size distribution of incipient flaws in a brittle material is generally assumed to follow a so-called Weibull power-law distribution characterized by two material parameters that indicate the number density of cracks that activate at or below a certain strain; the larger the body, the larger the biggest crack and the lower the strain at which it activates.[15] Consequently, the critical impact energy per unit target mass for disruption, denoted $Q^*$, decreases with the size of the parent body in the strength regime. However, the magnitude of this decrease with size is still unclear. At larger sizes, eventually a limit is reached at which gravity can no longer be neglected and the critical impact energy $Q^*$ starts to increase. This is the gravity regime.

The transition between the two regimes, as estimated by numerical simulations but still under investigation, occurs at sizes around 100 meters in diameter for typical rock. Moreover, in the gravity regime, as we will discuss below, the largest final fragment is generally larger than that produced solely by the impact. When fragments escape a large body, they can still be massive enough to be attracted by each other, and this can lead to some reaccumulation. Thus, at the end of the full event, most of the large fragments will consist of aggregates formed by reaccumulation, while in the strength regime all fragments are monolithic pieces. This result requires a distinction between two kinds of $Q^*$ impact energy thresholds: the first one is $Q^*_S$ and corresponds to the impact energy required to *shatter* a body so that the largest intact fragment produced contains 50% of the target's mass. The second is $Q^*_D$ and corresponds to the specific impact energy required to *disperse* the fragments so that the largest one contains 50% of the target's mass. In the strength regime, $Q^*_S = Q^*_D$ as all fragments are intact/monolithic. Conversely, in the gravity regime, since reaccumulation takes place, the largest intact fragment is always smaller than the largest fragment produced by reaccumulation, so that $Q^*_D > Q^*_S$ in this regime (see Fig. 2).

It is important to point out that these critical energies for a fixed target diameter also depend on the impact speed, impact angle, and on material properties. These dependences have been investigated in detail only for brittle non-porous materials; numerical models for porous materials are still being developed.[16,17] Similarly investigations of $Q^*$ for weak ice are underway, so far finding a slightly smaller $Q^*$ for the same impact speeds as those used for basalt or strong ice materials.[18] However, much work remains to be done to determine the sensitivity of $Q^*$ on the material parameters, especially porosity.

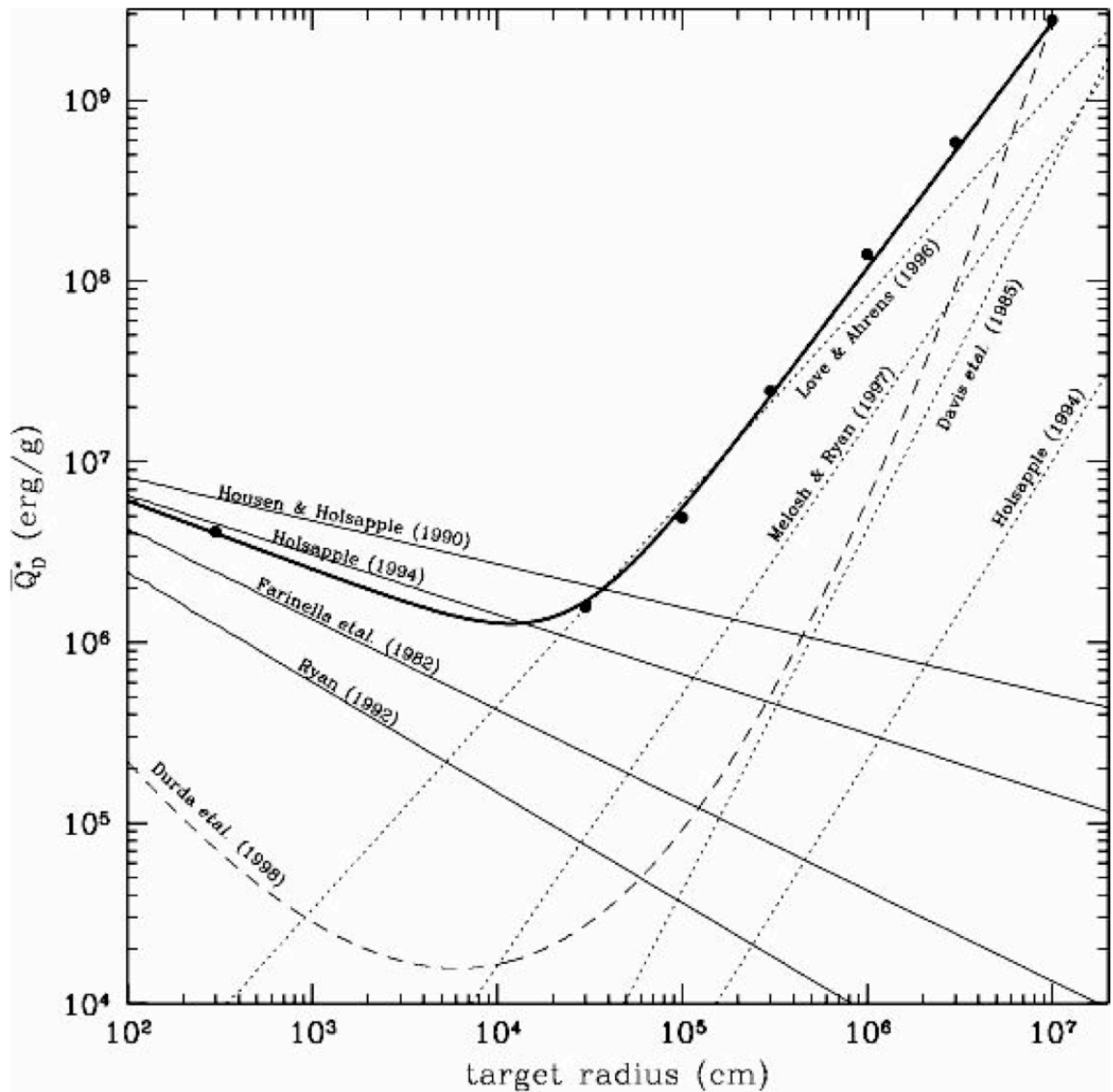

**Figure 2**: Impact energy required to disrupt a non-porous target such that the largest dispersed fragment contains 50% of the target's mass as a function of the target radius. The impact energy is defined as Q=E/M (erg/g), where E is the kinetic energy of the projectile and M is the target's mass. The impact speed for these curves is about 5 km/s, i.e. the typical impact speed between Main Belt asteroids.[3] The only measured values are those at laboratory scale around a few centimeters. At larger scales, the different curves correspond to the different studies labeling those curves. The bold solid line corresponds to estimates by [19] from numerical simulations using a three-dimensional smoothed particle dynamics (SPH) hydrocode. The large differences between these curves shows how poor our understanding of the process remains. Reprinted from [19] Icarus, 142, Catastrophic Disruptions Revisited, W. Benz and E. Asphaug, 5-20., Copyright (1999), with permission from Elsevier.

Theoretical models of catastrophic collisions attempt to bridge the gap between the very small experimental scales and those relevant to asteroids through non-dimensional relationships. These "scaling laws" account for projectile size, impact speed, target strength, density among a number of parameters and are necessarily idealized as they assume a uniformity of the process as well as a structural continuity and homogeneity.[20] Though they remain useful as a guide, they cannot predict with a high degree of reliability large-scale impact outcomes.

Numerical simulations remain the best approach for studying the collisional process at large scales. It is now possible to simulate complex disruptive impacts at high resolution thanks to the development of dedicated numerical codes and rapid improvement of computer performance. Impact experiments in the laboratory remain necessary to validate these models at small scales before they are applied to large-scale problems. In the 1990s, a hydrodynamical code (hydrocode) based on the Smoothed Particle Hydrodynamics (SPH) method was developed by,[21] and included a model of brittle failure. It has successfully reproduced the results of centimeter-scale laboratory experiments on basalt targets (this code is discussed in further detail below.) However, in the size range of asteroids (> 100 meters), its sole use is not sufficient due to the significant role that gravity can play in determining the final post-impact state. In a collision involving large bodies, ejected fragments produced by the fragmentation process will interact gravitationally. Therefore, reaccumulation can occur when relative speeds between fragments are below their mutual escape speed, leading to a distribution of large aggregates. Accounting only for the fragmentation phase and neglecting gravity could prevent the formation of such aggregates, missing an important part of the experiment result. A successful numerical approach has been to carry out the impact using SPH and then follow the subsequent gravitational interactions of the fragments using a gravity code (see e.g. [22,23]).

To determine whether both the understanding of the collisional process and the outcomes of the simulations are correct, we have at our disposal a unique laboratory at the appropriate scales: asteroid families. More than 30 asteroid families have been identified in the asteroid belt, each corresponding to a group of small bodies sharing the same spectral and dynamical properties (see e.g. [24-27]). The similarity of properties suggests that all members within a group once belonged to a larger asteroid, the parent body, which was catastrophically disrupted by an impact with a smaller projectile at high speed. Therefore, each asteroid family represents the outcome of a large-scale collisional event, and any numerical model should be able to reproduce its main characteristics.

The first numerical simulations that successfully reproduced these large-scale events,[22,23,28,29] found that when a large parent body (several tens of kilometers in diameter) is disrupted by a collision, the fragments interact gravitationally during their ejection, resulting in reaccumulation and the formation of aggregates. The final outcome of such a disruption is thus a distribution of fragments, most of the large ones being aggregates formed by gravitational reaccumulation of smaller ones. The implication of these results is that most large family members should be rubble piles and not monolithic bodies. Moreover, it was found that collisional disruptions naturally form binary systems and satellites, which have been found with properties similar to those predicted.[22,30]

The physics of the gravitational phase, where fragments evolve under their mutual attraction, relies on the fundamental laws of classical mechanics. However, there are aspects to the problem that make numerical simulations challenging. First, the number of generated fragments can number in the millions. Mutual gravitational calculations for millions of particles requires efficient numerical methods to save CPU time. Moreover, during their evolution, these fragments also undergo physical collisions, increasing complexity in the simulations.

A numerical *N*-body code, `pkdgrav`, was developed to compute the gravitational and collisional evolution of large numbers of particles and has since been applied to asteroid impact investigations.[31] During the gravitational phase of the simulations, collisions are typically assumed to not cause further fragmentation, but only result in merged particles or simple inelastic collisions.  This simplification is justified by the fact that the relative speeds between the ejected fragments are small enough that collisions between the pieces are quite gentle, with energies below any fragmentation threshold.  So, when collisions occur during this phase, depending on some speed and spin criteria, the particles either merge into a single particle whose mass is the sum of the particle masses and whose position is at the center of mass of the particles, or bounce with some coefficient of restitution to account for dissipation (see [28] for details).  In this approximation, the aggregates that are formed have "correct" masses, but their shapes are all spherical because of the merging procedure. Of course, the final shape and spin distributions are also an important outcome of a disruption and simulations have recently achieved a major improvement: instead of merging into a single spherical particle, colliding particles are able to stick together using rigid body approximations.  Such improvements will allow the determination of the shapes and spins of aggregates formed during a catastrophic disruption.[32]

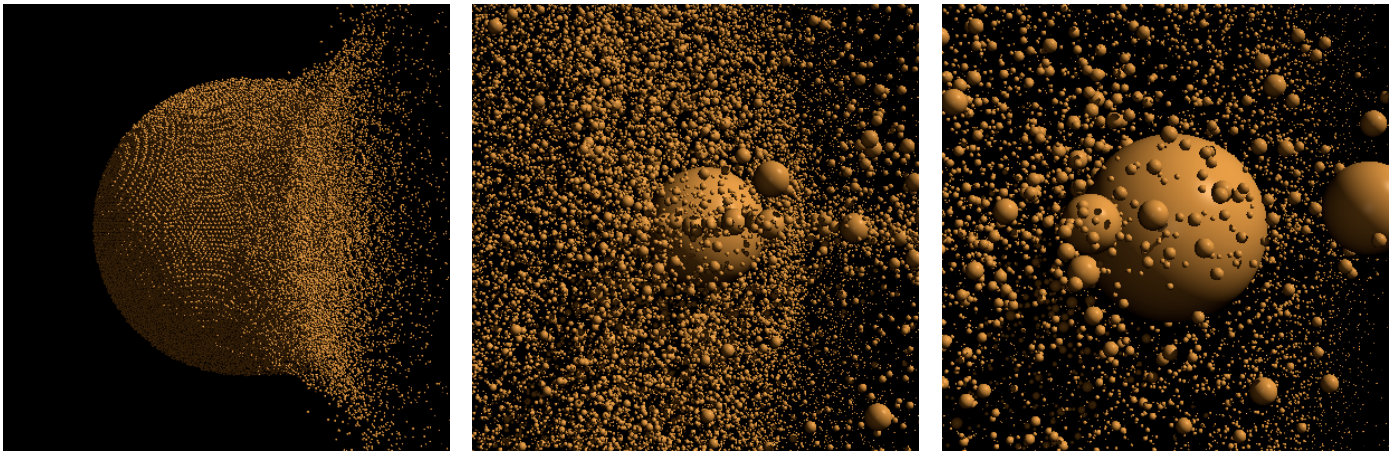

**Figure 3**: Gravitational phase of the collisional disruption of a 100 km-size asteroid. The first snapshot shows the instant after complete fragmentation of the asteroid.  All particles are a few tens of meters in size. After a few minutes, some of these particles start reaccumulating, as shown in the second snapshot.  After a few tens of minutes, up to a few days, the largest remnant shown in the last snapshot is formed. In fact, by the end of the simulation, all fragments, apart from the smallest ones, consist of gravitational aggregates.

The most poorly understood part of the collisional process is the fragmentation phase, immediately following the impact of the projectile.  It usually lasts twice the time for the shock wave to propagate through the whole target (a few seconds for a kilometer-size body).  The process of rock fragmentation is still a wide-open area of research, relying on a large number of assumptions based on a limited amount of data. A high-speed impact into a rock generates a shock wave, followed by a rarefaction wave, that initiates crack propagation. Thus, for the purpose of modeling a collision the rock can be seen as a continuum, but it also contains some discrete elements (the initial cracks).  This mixture of continuum and discrete features makes the development of a numerical scheme difficult.

A numerical code used to compute the fragmentation phase is generally called a hydrocode, which emphasizes the fact that this process involves the physics of hydrodynamics, although it occurs in a solid. The main difference between a fluid and a solid is that the deviatoric (non-diagonal) part of the stress tensor is not null in the case of a solid, while in a fluid only the spherical (diagonal) part of the stress tensor representing the pressure is important. Because of this difference there are three kinds of waves (elastic, plastic and shock) that propagate through rock during an impact. Elastic waves are well known and determined by linear relationships between the stress and strain tensors. Plastic waves begin to develop when the material strength changes with the wave amplitude. Then, at wave amplitudes that are high enough and associated with shock waves, the body is treated as a fluid.

Being non-linear, the transitory behavior between these types of waves is difficult to determine analytically from constitutive models, motivating the development of numerical algorithms. The process has been studied by implementing the bulk properties of a given rock in a numerical model of a continuous medium (a hydrocode), including a yielding criterion and an appropriate equation of state. The main advantage of this method is that no assumption on the form of the stress wave that drives the fragmentation is required. The initial conditions evolve numerically based on a rheological model and a failure criterion, and the appropriate regime (elastic, plastic, or shock) is determined by the computation.

The 3D Lagrangian hydrocode developed by Benz and Asphaug (1994) represents the state-of-the-art in numerical computation of dynamical fracture of brittle solids. Using SPH, the values of the different hydrodynamic quantities are known at discrete points (particles) and move with the flow (see [33] for a review of this method). Starting from a spatial distribution of the particles, the SPH technique allows the computation of the spatial derivatives without the necessity of an underlying grid. The 3D SPH hydrocode is thus able to simulate, consistently from statistical and hydrodynamical points of view, the fragments that are smaller or larger than the chosen resolution (total number of SPH particles). This approach successfully predicts the sizes, positions, and velocities of fragments measured in laboratory experiments for basalt targets, without requiring the adjustment of too many free parameters;[21] moreover, associated with the *N*-body code `pkdgrav`, it has successfully reproduced the main properties of S-type asteroid families produced by the disruption of parent bodies.[22,23,28,29]

Despite these recent successes of impact simulations, reproducing some experiments and asteroid family properties, there are still many issues and uncertainties in the treatment of the fragmentation phase. Some of the most important are:

- Material parameters: one of the main limitations of all research devoted to the fragmentation process comes from the uncertainties regarding the material properties of objects involved in the event. For example, ten material parameters describe the frequently adopted Tillotson equation of state (see [34], appendix II). Other sensitive material-dependent parameters are the shear and bulk modulus, but the most problematic parameters are probably the two Weibull parameters *m* and *k* used to characterize the distribution of initial cracks in the target. In fact, experimental data are quite scarce for these parameters due to the difficulty of determining their values. This is a critical problem because the validation of numerical simulations by comparison with experiments has been done by freely choosing those missing values so as to match the experiments.[21] This is not a completely satisfactory approach for an *ab initio* method such as the one provided by SPH simulations, though this is often the only option available. A database including both the material parameters of targets and outcomes of impact experiments using these targets is thus required to perform a full validation of numerical codes. Such a project has begun, using the experimental expertise of Japanese researchers from Kobe University and the numerical expertise of French and Swiss researchers from Côte d'Azur Observatory and Berne University. Early results include measurements of Weibull parameters of a Yakuno basalt used in impact experiments.[35]

- Crack propagation speed in a solid: the value of the crack growth speed is usually assumed to be 40% of the longitudinal sound speed in numerical simulations. This speed influences the number of cracks that can be activated for a given strain rate. Thus it plays a major role in the number and sizes of fragments that are eventually created. The lack of measurements of this speed and its possible dependence on material type leave researchers no choice but to use an intermediate value, such as the one currently assumed (about 40% of the sound speed of the material). However, it is important to keep in mind that this may need to be revised once some cracks are found to propagate at higher/lower speed in a sufficient number of experiments. The dynamical strength of a material does not depend only on its size but also on the applied strain rate, and the exact relation between strain rate and dynamical strength is a function of the crack propagation speed.

- Model of fragmentation: up to now, all published simulations of impact disruption have been done using the Grady-Kipp model of brittle failure.[36] In this model, damage increases as a result of crack activation, while microporosity (pore crushing, compaction) is not treated. However, several materials contain a certain degree of porosity (e.g. pumice, gypsum), and asteroids belonging to dark taxonomic types (e.g. C, D) are believed to contain a high level of microporosity. The behavior of a porous material subjected to an impact is likely to be different than the behavior of a non-porous one, as already indicated by some experiments (e.g. [37]). Therefore, a model for porous materials is required in order to address the problem of dark-type asteroid family formation and to characterize the impact response of porous bodies in general (including porous planetesimals during the phase of planetary growth). Such models have been developed recently and added to numerical codes (e.g. [38,16]). However, their application has been limited to cratering experiments and not all have been fully validated. Recently the so-called P-alpha model in their 3D SPH hydrocode and showed for the first time a successful comparison with impact experiments on pumice made at Kobe University.[16,17] Although further tests will be performed with other porous materials, this first validation will allow the use of their model at large scales with applications to the formation of dark type asteroid families, families in the Kuiper belt, and the general characterization of the catastrophic impact energy threshold ($Q^*_D$) for such bodies.

- Rotating targets: all simulations of catastrophic disruption have been performed starting with a non-rotating target. However, small bodies are spinning, and the effect of the rotation on the fragmentation is totally unknown. Some preliminary experiments have been performed suggesting that, everything else being equal, a rotating target is easier to disrupt than a non-rotating one (K. Housen, private communication). If confirmed, this will be an important result, as all models of collisional evolution of small body populations use prescriptions that are provided by numerical simulations on non-rotating targets. In particular, the lifetime of small bodies may be shorter than expected if their rotation has a substantial effect on their ability to survive collisions. It will thus be important to characterize the impact response of rotating bodies, both experimentally and numerically.

**Conclusion on the collisional process**

Our understanding of the collisional processes between small bodies of our Solar System has greatly improved in the last decade, thanks to the development of experiments, analytical theories, and sophisticated simulations. However, there are still many uncertainties and problems that need to be investigated. The outcome of some impact experiments have been reproduced successfully through the use of a 3D SPH hydrocode that incorporates the Grady-Kipp model of brittle failure. The inclusion of the gravitational phase using a sophisticated *N*-body code has also allowed for successful modeling of asteroid families caused by catastrophic collisions.

Collisional processes play a fundamental role in the different stages of the history of our Solar System, from planetary growth by collisional accretion to the current stage where small bodies are catastrophically disrupted. Moreover, the determination of the impact response of a small body as a function of its physical properties is crucial for the definition of efficient mitigation strategies aimed at deflecting a potentially threatening near-Earth asteroid whose trajectory leads it to the Earth. However, there are still many issues that need to be addressed to improve the accuracy of impact modeling. Detailed laboratory impact experiments need to be performed on a wide variety of materials in order to better understand the processes at small scale and to help validate the numerical codes, which can then be used to explore larger scales.

**2. Rotational Disruption**

**Introduction to Rotational Disruption:**

Rotational disruption of asteroids has emerged as an important process for altering individual bodies and shaping an entire population. Tidal disruption, a long-studied phenomenon, reminded the world of its importance when comet Shoemaker-Levy 9 was torn into at least 21 pieces during a close approach with Jupiter in 1992, and returned to collide with the planet two years later. The late 1990's brought about a spate of discoveries of satellites around asteroids, many of which hinted at an origin of rotational disruption. Rubincam (2000) explored a radiation effect acting on small bodies that could double or halve spin rates and dramatically shift asteroid obliquities on million-year timescales: far shorter timescales than collisions or planetary encounters.[39] This effect, called YORP,[39-42] operates due to the "windmill" effect of radiation reflection and blackbody emission from the surface of an irregularly shaped asteroid, and was immediately identified as having the potential to spin asteroids so fast that they could distort or disrupt if they had little to no cohesion. Combined with a growing inventory of binary asteroids with rapidly rotating primaries and close secondaries, the YORP effect has helped bring greater attention to the profound effect rotational disruption has played in evolving the properties of small asteroids.

An important aspect of asteroid studies that has developed in the last 10–15 years is the realization that many, perhaps most, asteroids are likely "rubble piles" (or, more precisely, gravitational aggregates). Briefly, this evidence includes: observed spin rates of asteroids that do not exceed a critical limit governed strictly by the body's gravity; surprisingly low densities of asteroids suggesting large void spaces, as determined from the population of binaries and measurements by space missions; crater chains on the Moon and the Galilean moons suggesting a history of tidal disruption in the Solar System; and very large craters found on some asteroids that suggest an ability to absorb impact energy without global destruction as would be expected for a monolithic target.[44] Recently, the visit of the Japanese Hayabusa spacecraft to the near-Earth Asteroid Itokawa added to the growing evidence for rubble piles, even for asteroids as small as a few hundreds of meters. What Hayabusa found was an asteroid seemingly turned inside out, covered with large boulders and rubbly terrain devoid of any obvious craters. The observations made by the spacecraft strongly suggest Itokawa has a fairly low bulk density (about 1.9 g cm$^{-3}$) and a high porosity (up to 40%), with plenty of void space between possibly coherent pieces in its interior (see Fig. 4, [45]). Note that this kind of porosity, characterized by the presence of macroscopic voids (called macroporosity) must be distinguished from microporosity, characterized by the presence of tiny pores in the material, as found in carbonaceous meteorites.

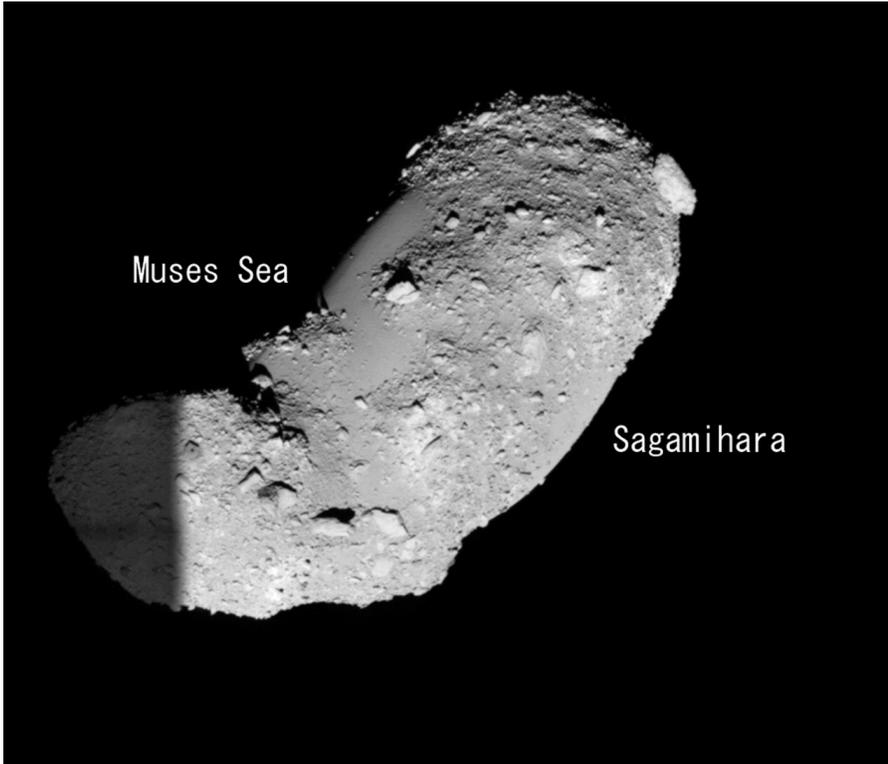

**Figure 4:** Photo of Itokawa taken from the Hayabusa spacecraft during its mission in 2005. The ~500 m asteroid shows a very rough surface covered by large boulders and a few smooth areas. There is a near absence of craters. From [45], A. Fujiwara and 21 coauthors, Science, 312, 1330 **(2006)**. Reprinted with permission from AAAS.

Gravitational aggregates (of which a rubble pile is a special case with zero cohesion) behave differently under rotational or disruptive stress compared to a monolithic asteroid and thus need to be modeled using different techniques. As discussed above, collisional disruptions of monolithic bodies are typically modeled using hydrodynamic codes, as the simulations mostly involve impulsive short-term events, such as hyper-speed collisions. A body constructed of coherent pieces, held together by gravity and/or very weak cohesion, can be more sensitive to less impulsive and more gentle physical processes (i.e. tidal encounters, rotation rate increase), and can be modeled more conveniently with a gravity code that includes some form of repulsion or solid body collisions to represent mechanical interactions between pieces. Models for simulating rubble piles with rigid, colliding particles (spheres) have progressed rapidly, and have been applied to both tidal disruption and YORP spinup of asteroids.[46,47,48] In the following two sections we outline the theories, which drive the models, and discuss where the models stand now.

## A. Tidal Disruption

Tidal disruption is the process of re-shaping or tearing apart a body due to the differential gravity of a massive nearby body (typically a planet in the cases we will focus on). Across the length of the body, the gravity on the side nearest the planet experiences an acceleration greater than that of the far side. In a frame located at the center of mass of the body being disrupted, the effect is perceived as forces pulling outward along the line pointing to the planet. For the case of a rock attempting to remain on the surface of a rigid asteroid, it is simply a competition between the body's effective gravity (including spin and orbital effects) and that of the planet. When considering the distortion and disruption of the entire body, the competition is between the body's effective gravity, a failure criterion depending on the angle of friction of the material constituting the asteroid, and the planet's gravity. A common measure of the strength of the tidal force is the Roche limit, which is the limiting distance at which a synchronously rotating (same side always facing the planet) fluid body can approach a planet while still maintaining an equilibrium shape.[49]

The physics of a close flyby is more complicated: the encounter is quasi-impulsive and the planet exerts a torque on the disrupting body. As it passes by the planet on a parabolic or hyperbolic trajectory (not orbiting, just flying by), it experiences stretching as the tidal force of the planet strengthens, but also is torqued as the now-elongated body tries to keep its (now-growing) long axis aligned with the planet. Richardson et al. (1998) covered a large parameter space investigating cohesionless rubble pile flyby dynamics, and defined three classes of disruption that will be useful to keep in mind as we review work on tidal disruption (see Fig. 5):[46]

a.) Catastrophic or "SL9-type", **S-class**: An S-class disruption results in the largest remnant having less than 50% of the original progenitor's mass. This disruption is the most dramatic because the body is stretched into a long line of material by the tidal forces as it passes the planet. The stretching continues as the body recedes from the planet, with clumps forming out of the line of debris. The namesake of this class, comet Shoemaker-Levy 9, provides the perfect example, as it formed at least 21 fragments in a "string of pearls."
b.) Rotational breakup, **B-class**: This class involves disruptions where the largest remnant maintains 50–90% of the original mass. This more moderate disruption also begins with the body being stretched into a cigar shape, on a line towards the planet. An overall torque is produced on the elongated body, and centrifugal force partially overcomes the body's self-gravity, resulting in substantial mass loss.
c.) Mild disruption, **M-class**: The mild disruptions apply to those cases with less than 10% mass loss. These are similar to the **B-class**, with the initial body getting stretched and torqued, just to a less extreme degree, with only very small clumps, or individual particles escaping off the ends of the body. This case frequently ends with a highly elongated main body.

| S-class | B-class | M-class |
|---|---|---|
| 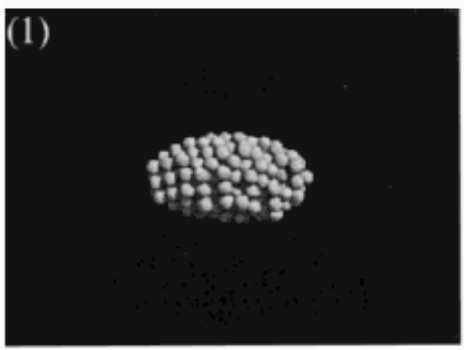 | 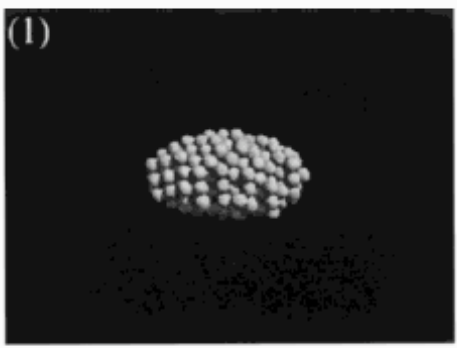 | 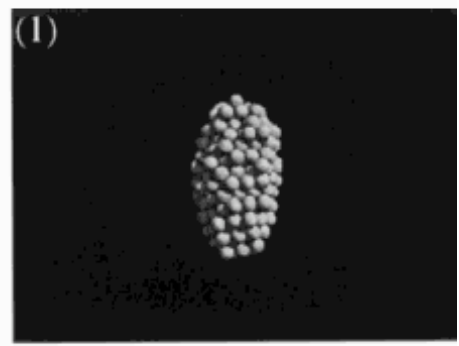 |
| (1) | (1) | (1) |
| 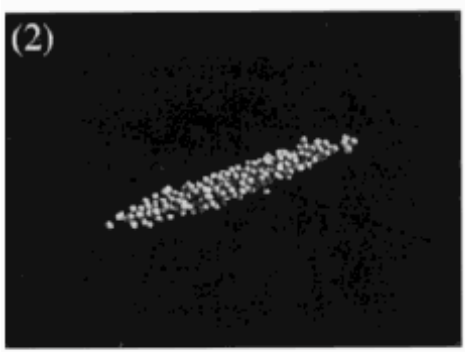 | 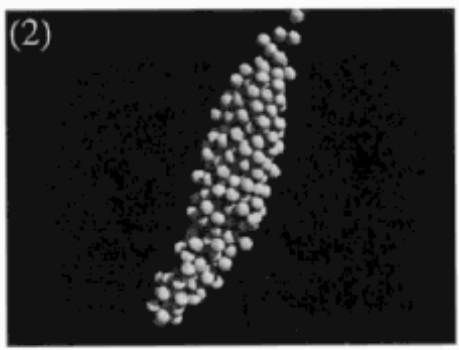 | 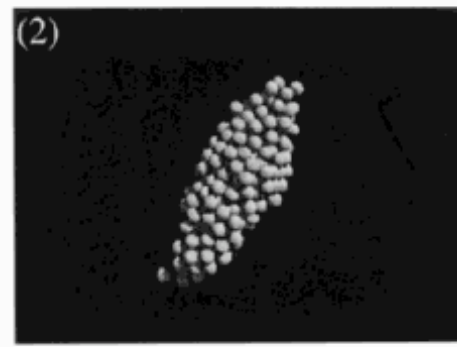 |
| (2) | (2) | (2) |
| 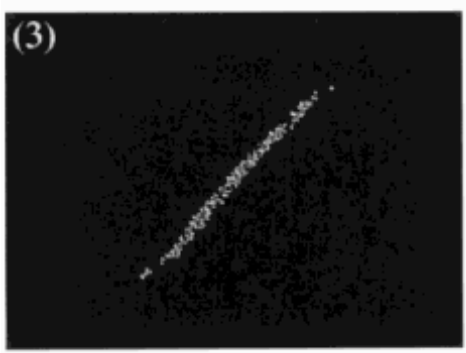 | 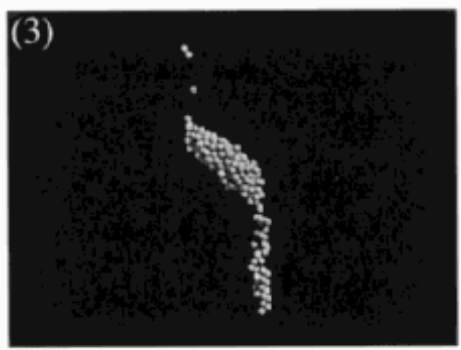 | 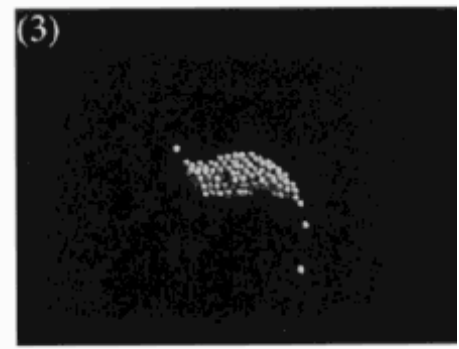 |
| (3) | (3) | (3) |
| 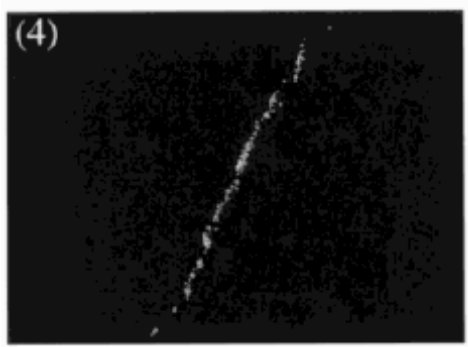 | 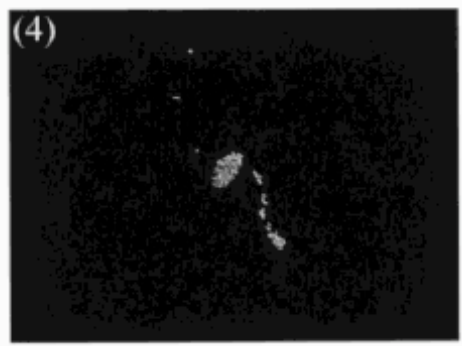 | 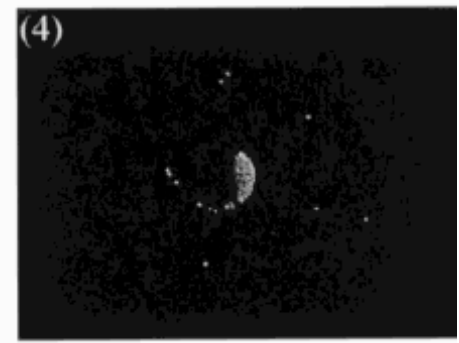 |
| (4) | (4) | (4) |
| 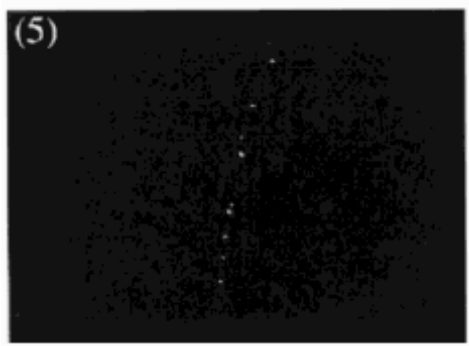 | 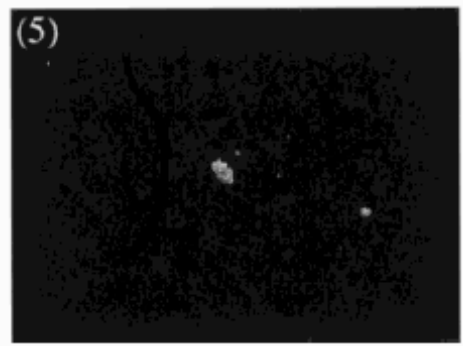 | 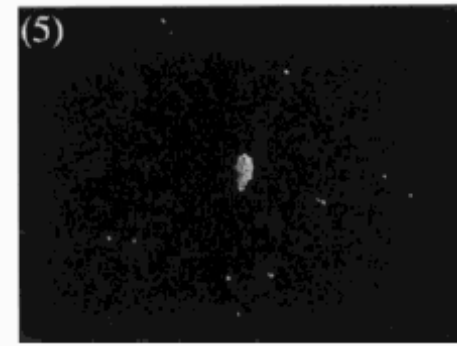 |
| (5) | (5) | (5) |

**Figure 5**: Snapshots illustrating the three classes of tidal disruption as defined by Richardson et al. (1998). Each column represents a different class of disruption, left is S-class, middle is B-class and the right column is the M-class. For each column, time proceeds from top to bottom (with each frame selected to show the progression, and not evenly spaced in time). Reprinted from [46] Icarus, 134, Tidal Distortion and Disruption of Earth-Crossing Asteroids, D. C. Richardson, W. F. Bottke and S. G. Love, 47-76., Copyright (1998), with permission from Elsevier.

The classes provide a qualitative and quantitative way to approach the wide range of tidal disruption outcomes. Before the assistance of numerical simulations made visualization and experimentation possible, a range of analytical estimates for disruption limits were derived for varying types of encounters and body structures. These are discussed briefly in the following section.

i. **Analytical Formulations**

Tidal disruption was first investigated analytically by Roche in 1847 (see [50]). It was shown that a self-gravitating liquid satellite, synchronously rotating, on a circular orbit around a spherical planet, has no stable equilibrium shape inside a critical distance. The synchronous rotation ensures that the same face of the body is oriented toward the planet throughout its orbit, and only adds minimally to disruption by way of centrifugal acceleration as the orbital periods are long compared to critical rates. This oft-quoted Roche limit, $r_{Roche}$, is

$$r_{Roche} = 1.52 \, (M_{pl}/\rho_{sat})^{1/3} = 2.46 \, R_{pl} \, (\rho_{pl}/\rho_{sat})^{1/3}$$

where $M_{pl}$, $R_{pl}$, and $\rho_{pl}$ are the mass, radius and density, respectively of the planet, and $\rho_{sat}$ is the density of the satellite. For example, a fluid body with density of 2 g cm$^{-3}$ would disrupt if orbiting closer than 3.4 Earth radii. Though the Roche criterion is a special case (fluid body, synchronous rotator, circular orbit), it is a standard metric and is used frequently in planetary science to estimate regions, for example, where satellites could, or could not, accumulate in a ring for a given density, or where satellites could start to disrupt.

More applicable to a body having a close-encounter with, rather than orbiting, a planet has been investigated for viscous self-gravitating bodies on parabolic encounters with planets.[51] This is a semi-analytic formalism and determined a periapse distance, $r_{disrupt}$, at which mass-shedding would occur,

$$r_{disrupt} = 0.69 \, r_{Roche} = 1.05 \, (M_{pl}/\rho_{pro})^{1/3} = 1.69 \, R_{pl} \, (\rho_{pl}/\rho_{pro})^{1/3},$$

where $\rho_{pro}$ is density of the passing body. This prediction is in good agreement with contemporary modeling work, showing that SL9-type disruptions are possible, and also finding that a flyby must come much closer than the Roche limit to instigate a mass-shedding disruption.

Tidal disruption from a different perspective, modeling the disrupting bodies using a cohesionless Druker-Prager strength criterion.[52] This model aims to capture the physics of dry granular materials, such as sand or piles of rocks on Earth, or rubble pile asteroids (see Spin-up section below for further discussion). This approach allowed for tidal disruption limits to be estimated for a range of potential body structures, with the material parameter, angle of friction, defining anything from a perfect fluid to standard terrestrial material. This treatment recovered the semi-analytic results above and made predictions about potentially hazardous asteroid 99942 Apophis and its future close approach. Similar results were obtained by modeling an asteroid as it transitions from a rigid body to a granular body using comparable strength models to determine failure, but volume-averaged.[53] The Druker-Prager treatment was extended to include limits for bodies with cohesive strength.[54] It was found that cohesive bodies can approach much closer to a planet and adopt a wide range of shapes, explaining the presence of small Saturnian satellites at distances to Saturn much smaller than the classical Roche limit and putting some constraints on their material properties.

### ii Numerical Models

The analytic formulations above provide mass-loss predictions for various types of bodies and materials under rotational stress. To determine what happens after the disruption, however, generally requires a numerical approach. Most numerical models to date have consisted of $N$-body simulations of self-gravitating rubble piles. The vital aspects are a model capable of combining self-gravity and particle collisions, and the capability to run fast enough to explore a wide parameter space.

Tidal disruption has been invoked to explain the disruption of Comet D/Shoemaker–Levy 9 (SL9). SL9 disrupted when the comet passed within 1.36 $R_J$ of Jupiter on 1992 July 7. The comet was torn apart and 21 fragments or reaccumulated clumps were later observed. $N$-body studies have since matched many of the comet's basic post-disruption features (train length, position angle, and morphology) using a strengthless rubble pile model.[55-58] Immediately after the discovery of the disrupted comet, estimates of its density were produced from the analytical studies detailed above, as well as numerical simulations. An upper bound on the density of the SL9 parent body was made using SPH simulations tidal disruptions, with the critical density, $\rho_c$, derived from those studies,

$$\rho_c < 0.702 \, (1.62 \, R_{Jup}/r_p)^3$$

based on the breakup limit,

$$r_p = 1.31 \, R_{pl} \, (\rho_{pl}/\rho_{pro})^{1/3}$$

where $r_p$ is the perijove distance, at the time the perijove was assumed to be 1.62 $R_{Jup}$, which required the density to be less then 0.702 g cm$^{-3}$ (the close approach distance would later be refined to ~1.3 $R_{Jup}$, which would suggest a maximum density of ~1.3 g cm$^{-3}$).[59,60]

As we have seen, the disruption limit for a body is dependent on the ratio of the density of the target to the planet, but also on the geometry of the encounter. Since planetary densities are precisely known, the density of the disrupting body can be estimated if a formulation for the correct encounter geometry can be determined. As shown in Fig. 5, there is a wide range of possible disruption outcomes, so an observed disruption does not necessarily equate to a density value, but rather places an upper limit. Detailed determination of density requires morphological matches between numerical simulations and the observed disruption.

Attempts at matching the morphology of SL9's breakup using *N*-body simulations first focused on the manner in which it coalesces into fragments after the tidal disruption.[61] The models consisted of a spherical collection of 321 identical particles that interacted gravitationally and with non-adhesive frictionless scattering to model collisions with some kinetic energy loss based on a simple coefficient of restitution. A density range between 0.5 and 0.6 g cm$^{-3}$ was determined from morphological matches of the fragment train formation for a close approach of 1.36 $R_{\text{Jup}}$. Rotation rate was examined qualitatively, but it was noted that varying rotation rate and direction had a measurable effect on both the resultant position angle of the fragment train and the morphological properties of the coalescing fragments.

The *N*-body models of SL9 increased in complexity by employing a model of hard spherical particles held together by their self-gravity, and kept apart by a restoring force that activated whenever particles overlapped.[56,57] The model comets were constructed of a varying number of particles, from 21 to 5000, and densities ranging from 0.1 to 2.0 g cm$^{-1}$, and sent the rubble piles on a flyby of Jupiter. Using the length of the fragment chain, and its qualitative morphology, they were able to estimate a diameter and density for the comet. This work was expanded, focusing on more general applications of tidal disruption and setting limits for different magnitudes of disruptions for varying densities, as well as studying some basic effects of rotation.[56]

The best density determination was tempered by the degeneracy between density and rotation rate. Simulations showed that even a moderate 9 h rotation of the comet with a 1 g cm$^{-3}$ bulk density gave nearly identical post-disruption morphology as the best fit with no rotation.[56] Tests with retrograde rotation resisted breakup. While the fragment train morphology was used to fit density/rotation, the length of the train was diagnostic of the size of the comet. This was used as a final parameter: after the morphology was matched, the comet diameter was scaled to match the train of fragments. Further attempts uses the detailed record of the evolution of the fragment train position angle on the sky as a diagnostic of the density/rotation degeneracy.[58] Thanks to improved computational capabilities, these simulations covered a wider range of parameters (including body orientation) prior to encounter. They found only minor variation in the position angle as a function of spin-axis orientation, and were not able to improve on the density/spin estimates.

The most general *N*-body study of tidal disruption was that for NEAs, covering a large parameter space of elongated rotating bodies (constructed with 247 particles) passing Earth at various close-approach distances *q* and encounter speeds $v_\infty$ (Fig. 5).[46] The study was designed to quantify disruption and mass loss for tidal encounters, but also tracked mass loss in an attempt to characterize binary asteroid formation. The simulations sparsely covered the parameters of *q*, $v_\infty$, progenitor elongation, progenitor spin rate, and long axis alignment. Basic trends in disruption were observed, with increasing disruption for closer approaches, slower approach speeds, and faster prograde rotation rates. More subtle results were seen as a function of body elongation and long axis alignment at close approach.

Significant upgrades to the numerical codes and computing power allowed a more general exploration of the large number of parameters governing a tidal disruption outcome. The numerical code, `pkdgrav`, was originally built for use in cosmological simulations and reduced computation time by using a tree to speed up gravitational calculations. A variant of the tree is used to search efficiently for particle collisions in the correct time order. Tidal disruption was re-visited using `pkdgrav` in an even larger suite of simulations at higher resolution, ~1000 particles in each body, designed to characterize binary systems that were created.[47] The parameter space covered was similar to that previously covered, but with progenitor spin axes randomly oriented over all possible values, instead of selected from a limited sample. This meant that for each encounter distance and speed, 100 random spin-axis orientations were considered. This work found substantial binary formation, with elongated primaries, and eccentric binary orbits. Though these binary systems are not likely a good match for most of those observed among NEAs (see next section), the large study of tidal disruption, and the new approach to studying the effect of the spin axis orientation on the outcome, provided some new insights into the mechanism as a way to disrupt NEAs. Primarily, spin axis orientation is very important in determining whether there is any disruption, or, if there is, whether it is substantial. Spin axis orientation also affected the final obliquity of the primary and inclination of the secondary orbit, allowing for minor (~5 degree) variations from the plane of the flyby.

A less dramatic effect of a tidal encounter is simply re-shaping or distorting of a body. The stretching and torquing of a rubble pile can leave a distinct shape even without mass loss. Material flows to fill the equipotential surface from the tidal forces, but the torque from the flyby can twist the body enough to create an S-shape or cusped ends. Maximum elongations were tested using *N*-body simulations and compared them with the extremely elongated asteroid Geographos.[61] This approach was taken a step further by comparing tidal distortion simulations with radar observations of the same asteroid.[62] The well-defined shape of Geographos matched many features of the simulation output, including the cusped ends, an opposed convex side, and a nearly concave side with a large hump. Thus tidal encounters may be at the origin of some of the unusual asteroid shapes that have been observed.

## B. Spin-up

### i. Equilibrium shapes for rotating bodies

Disruption by *in-situ* spin-up has only recently become a significant field of research concerning asteroids. Previously, investigations of equilibrium/spin-limit shapes were restricted to self-gravitating fluids/clouds rather than self-gravitating rubble piles (see [50]). However, with the renewed interest in thermal effects on km-sized asteroids, equilibrium shapes and responses to spin state changes have become recognized as important in understanding many phenomena affecting small asteroids in the Solar System. As well, rotational disruption was largely suspected to be the cause of binarity among small asteroids.[63]

A particle on the surface of a rotating body feels an acceleration from the gravitational force of the body directed inward and from the centrifugal acceleration due to the body's rotation directed outward. A crude spin limit for material to remain on the equator is obtained by balancing these opposing forces. For a rigid sphere and a test particle at the equator, this balance is $Gm/r^2 = \omega_c^2 r$, where $\omega_c$ is the critical angular speed of rotation, $G$ is the gravitational constant, $m$ the mass of the sphere, and $r$ is its radius. The corresponding critical period $P_c = (3.3hrs)/\sqrt{\rho}$, where the density $\rho$ is expressed in g cm$^{-3}$. If the primary body is not homogeneous or spherical, the calculation becomes dependent on the exact shape and distribution of the mass internally. The simplest extension of this previous calculation is for a particle on the long end of a proloid, where the ellipsoid's gravity field is approximated as being that due to a point mass, giving the adjusted limit $P_c = (3.3hrs/\sqrt{\rho}) \times \sqrt{(a/b)}$, where *a* and *b* are the long and intermediate axis lengths (for full derivations of limits involving rigid proloids see [64]).

The search for equilibrium shapes of fluid bodies began with Newton (*Philosophiae Naturalis Principia Mathematica*)[65] who realized that the rotation of the Earth would force it to be an oblate spheroid, rather than a perfect sphere. An extension of this work by MacLaurin resulted in a class of equilibrium oblate spheroid shapes for self-gravitating fluids with larger ellipticity. Jacobi extended this work, finding a series of tri-axial equilibrium shapes closely related to the MacLaurin ellipsoids (see [66]). Chandrasekhar recovered all of these equilibrium figures in his comprehensive work on elliptical equilibrium figures (see Fig. 6).[50]

The classical Jacobi and MacLaurin ellipsoids define equilibrium spin and shape combinations, and though they rarely exactly equate to asteroid shapes, they are commonly used to compare with measured shape/spin combinations. The MacLaurin ellipsoids are oblate, meaning that the long and intermediate axes are of equal length, with the short axis always being shorter. The short axis of MacLaurin ellipsoids gets proportionally smaller as rotation rate increases, until a bifurcation point is reached, at which point the body transitions into a Jacobi ellipsoid with all three axes of differing lengths.

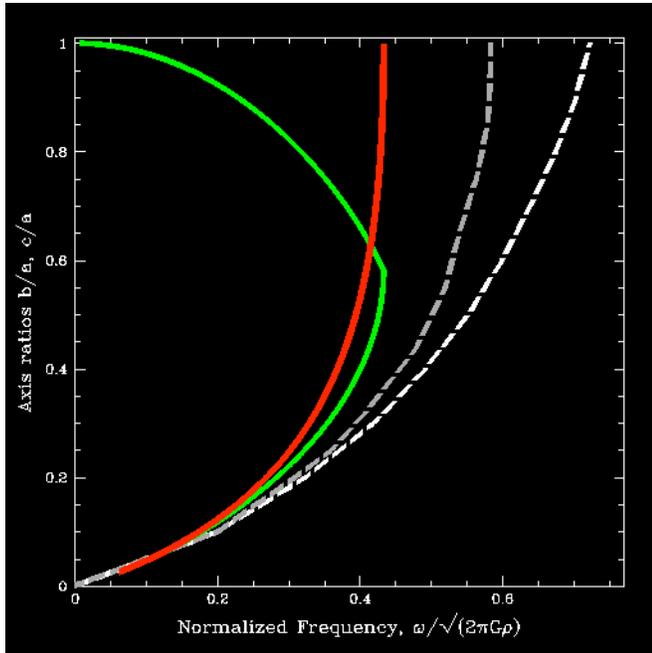

**Figure 6**: Classic fiducials and limits for equilibrium axis ratios as a function of the normalized angular spin frequency. The red and green lines represent the b/a and c/a axis ratios respectively for the classic Jacobi and MacLaurin shapes. The MacLaurin ellipsoids correspond to b/a = 1 and the upper branch of the green line, while the Jacobi ellipsoids are described by the lower branches of the green line and the entirety of the red line. The two dashed lines represent the boundaries for cohesionless granular material with internal angle of friction ϕ = 20° (left) and ϕ = 40° (right).

Shifting from perfect fluid formulations, theory for spin and shape configurations controlled by the Mohr-Coulomb friction angle $\varphi$ of the constituent material was developed.[67] The friction angle parameter is a guide to the shear strength of a material under pressure; for example, a perfect fluid has $\varphi = 0°$, while typical terrestrial materials, like sand, have $\phi \sim 30°$. The distinction is that a fluid never has shear strength, while sand can form a pile or resist shearing (quite simply, this is why we can walk on sand and not on water). The spin and shape configurations determined using this method match the classical Jacobi ellipsoids for $\phi = 0°$, while at moderate friction angles there is a region, or envelope, of allowable configurations centered around the fluid limits. This was then compared these allowable shapes to those for 845 asteroids, finding that nearly all asteroids are consistent with cohesionless structures, with moderate porosity.[67] This work directly related to the larger work on the rubble pile structure of asteroids, simply showing that this sample of asteroids *could* all be cohesionless bodies behaving according to the physics of granular material.

A similar study for "idealized" rubble piles (made up of identical, indestructible spherical particles), using `pkdgrav`.[64] This work differed from previous efforts in that the bodies were constructed of finite-sized particles, rather than being modeled as an analytical continuum, and in that the disruption could be followed to late time (whereas the analytical methods cannot proceed past the first instance of mass loss). A parameter space of various shapes and (fast) starting spins was explored. The bodies that started outside the limits for $\phi \sim 40°$, as defined by the Holsapple (2001) cohesionless limits, underwent re-shaping or mass loss, while the bodies starting inside the limit did not lose mass, despite often changing in shape or spin. The extreme mass-loss cases, where more than 10% of the mass was lost, generally re-shaped and ended up on, or near, the classical Jacobi ellipsoid limits. This means that such idealized rubble piles can be spun-up to the $\phi \sim 40°$ limit, but if disrupted will reaccumulate back to classical fluid shapes. This work linked the *N*-body rubble pile simulations with analytical determinations for expected asteroid shape and spin behavior, allowing more detailed work on how asteroids might respond to potential spin-state changes, particularly spin-up by the YORP effect.

### ii. YORP

Spin rates and obliquities of asteroids are subject to change due to the reflection and emission of radiation from irregularly shaped asteroids.[39] The crude model in Fig. 7 shows, by way of simple wedges added onto a perfectly symmetric spherical body, that a net torque is produced around the body's spin axis. The asymmetries found on a real asteroid arise from the large-scale shape of the body or smaller-scale orientations of surface elements. Thus a body reflects sunlight or re-emits absorbed energy anisotropically, creating net torques around the rotation axis. An asteroid covered by a layer of regolith absorbs and re-emits radiation with a different efficiency compared to monolithic rock; this property, thermal conductivity, determines how the body is heated internally and governs the temperature at its surface and its blackbody emission. YORP occurs regardless of the time of re-emission: it is simply dependent on the direction of emission, thus allowing the effect to work regardless of the thermal inertia of the body. This is important since determining detailed thermophysical models for asteroids is a difficult challenge. Also important is that the effect occurs for both reflected and re-emitted radiation, so even very dark asteroids will be affected.

The first analytical investigations of the YORP-effect ignored thermal conductivity out of simplicity. Fundamentally, thermal conductivity is not necessary for YORP to function, unlike the related Yarkovsky effect that depends on a time lag for peak thermal emission. Though YORP works with or without thermal conductivity, its overall effect will change with the addition of this parameter. The inclusion of thermal conductivity increases the potential change in obliquity and drives asteroid spin axes to states perpendicular to the orbital plane.[68] As well, the asteroids are equally likely to have their rotation rates accelerated or decelerated.[68]

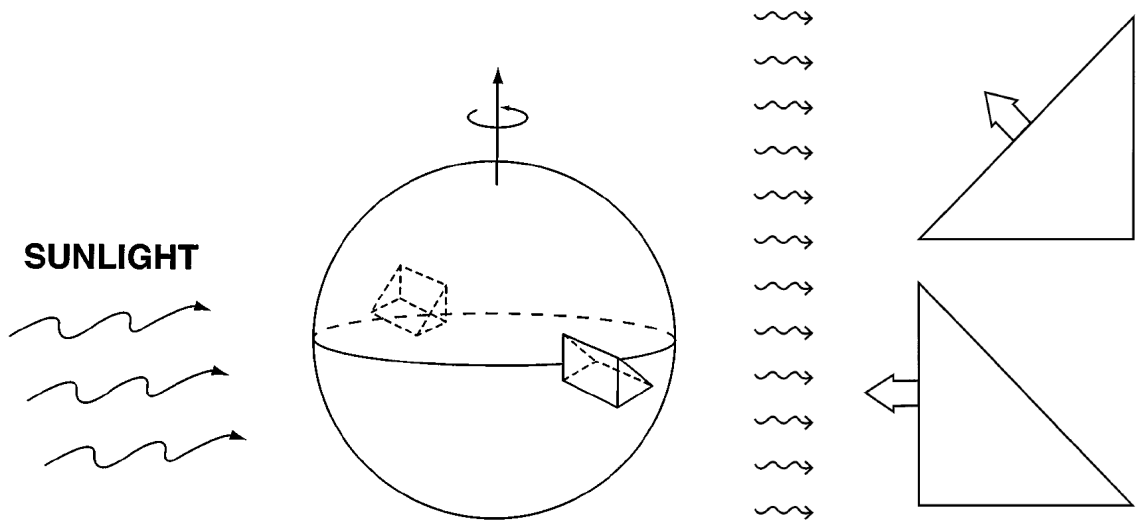

**Figure 7**: Left: A simple model of an asymmetric asteroid. The asteroid is symmetric except for the two wedges on its surface. Reflection of solar radiation, or blackbody emission directly from the body itself, causes the body to experience a net torque. Right: The two wedges on the asteroid have identical surface areas for absorbing radiation, but upon re-emission, according to Lambert's law, the net momentum of the photons leaving the surface is normal to the surface. Hence the wedge feels a push in the opposite direction. The force is the same, but in different directions, leading to the net torque about the rotation axis. Reprinted from [39] Icarus, 148, Radiative Spin-up and Spin-down of Small Asteroids, D. P. Rubincam, 2-11., Copyright (2000), with permission from Elsevier.

The first detection of the YORP effect was an elegant description of obliquity change based on a long-term series of observations determining spin-axis orientation and rotation periods for a subset of a very old asteroid family.[69,70] Following a catastrophic collision that created the Koronis family many billions of years ago, it was expected that the spin axes of the family members would be re-oriented randomly due to collisions, with spin rates filling in a Maxwellian distribution. When observed, this subset of large family members ($D < 40$ km) was found to have two distinct populations: prograde rotators with rotation periods between 7.5-9.5 hrs and obliquities between 42 and 50 degrees, and retrograde rotators with obliquities between 154 and 169 degrees, and either fast (< 5 hr) or slow (> 13 hr) rotation rates. A model of the YORP effect on spin rate and obliquity evolution tracked the changes in the spin states, and matched the observed parameters of the Koronis family. The prograde rotators were driven into a spin-orbit resonance, pushing them to their uniform properties, and the retrograde rotators were driven to extreme obliquity and either rapid or slow rotation.

The most direct observation of YORP acting on a single body is obtained by measuring a systematic change in a body's rotation rate over time. The obvious difficulty with this approach is that the timescale of YORP spin up or spin down is quite long compared to the baseline of most sets of observations. Nonetheless, instrument sensitivity is sufficient that this measurement has been made successfully for two different asteroids (1862 Apollo and 2000 PH5) using a suite of photometric lightcurves and radar observations (Fig. 8).[71-73] Both studies created shape models of the asteroid and used them to model the lightcurves over long baselines in order to make measurements of sufficient accuracy of rotational phase. All of these studies found measured YORP spinup near the calculated estimates based on a detailed model of each asteroid's shape and size.

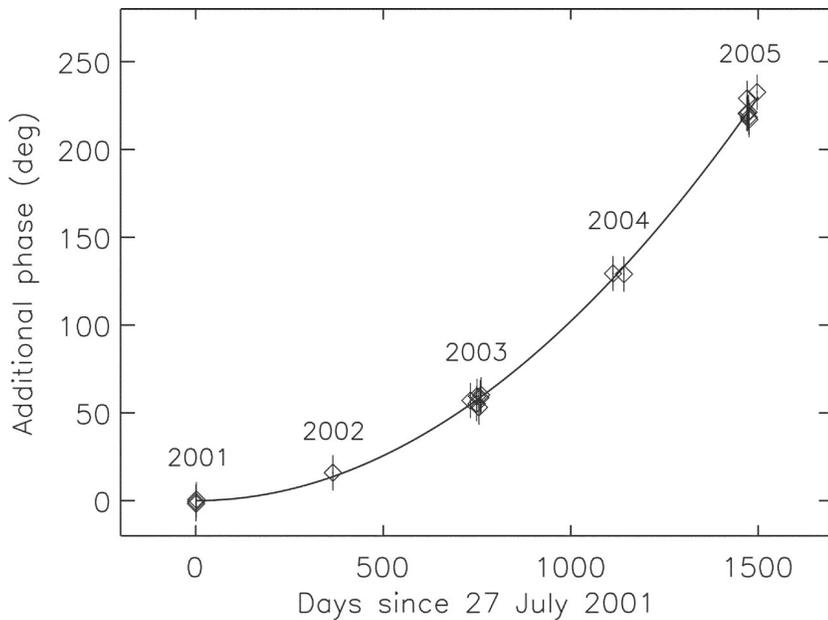

**Figure 8:** Rotational phase changes in asteroid 2000 PH5 plotted over time, showing the quadratic increase in time of additional phase. Lightcurves from 2001-2005 were combined with a radar shape model to determine the phases. Reprinted with permission from Taylor et al. 2007. From [71], P. A. Taylor and 11 coauthors, Science. 316, 274 **(2007)**. Reprinted with permission from AAAS.

Spinup of rubble pile asteroids was modeled generically by gradual increases in spin rate, mimicking the YORP-effect acting on a cohesionless asteroid.[48] The spin-up procedure was a simple series of discrete angular momentum boosts to the largest mass in the system, thus allowing any escaped particles to accumulate into satellites. The perfect monodisperse (same-size particle) rubble piles used with `pkdgrav` behave similarly to cohesionless granular material with an angle of friction ~40°.[64] This value is roughly what is expected for the behavior of asteroidal material, and is in line with typical terrestrial material. But the spinup models also modeled rubble piles with a bimodal size distribution of particles that, under spin tests similar to those previously used, behaved like material with different angles of friction. By changing the size ratio in the particle size distribution they created rubble piles with behavior matching a ~20° and ~0° angle of friction. By not having a strictly monodisperse population of particle sizes, the fundamental crystalline structure of closely packed spheres was perturbed, changing the way the bodies responded to spin-up.

As it was spun up, the fluid case, $\phi$ ~0°, behaved in a manner consistent with theoretical expectations for a fluid as established by the MacLaurin and Jacobi limits, with the body starting oblate, becoming prolate, and staying in a tri-axial prolate state for the extent of the simulations. These bodies lost mass from the ends of their long axis, and did not accumulate a satellite. Test bodies for the intermediate case, $\phi$ ~20°, also became tri-axial, but with less total elongation, and in some cases were able to accumulate a satellite from the orbiting debris. The nominal case with $\phi$ ~40° produced bodies that remained oblate for the entirety of the simulations, and accumulated satellites efficiently and consistently. In these cases a ridge of material formed at the equator, which was the zone of mass loss. When the particles on the initial rubble pile were color-coded and tracked for the duration of the simulation, a trend of clearing material from the poles to the equator was found, and the satellites were made largely of this original surface material (see Fig. 9).

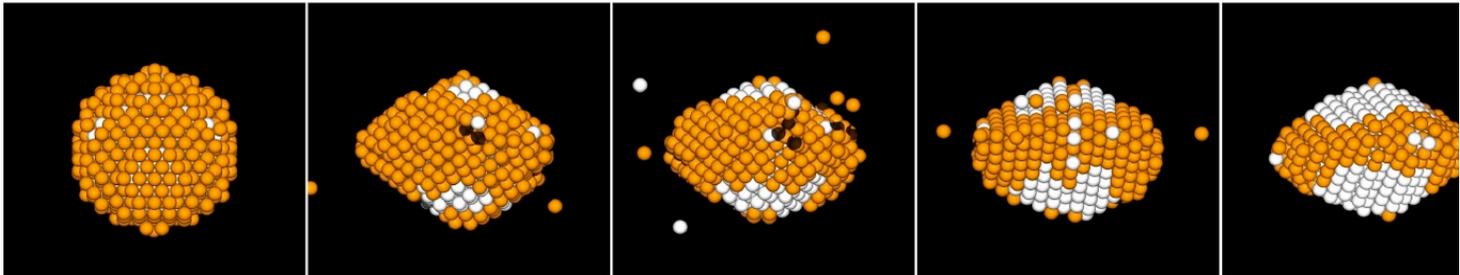

**Figure 9:** Figure from Walsh et al. (2008) showing the evolution of primary shape during YORP spinup. In this figure, the orange particles are those that start initially on the surface of the asteroid, while the white particles start in the interior of the model asteroid. The ridge at the equator of the asteroid is evident by the final frame, as is the clearing of original surface material, exposing many regions of white particles. The accumulated satellite is not shown. Reprinted by permission from Macmillan Publishers Ltd: Nature, [48], K. J. Walsh, D. C. Richardson and P. Michel, Nature. 454, 188 **(2008)**, copyright (2008).

Finally, the role played by contact binaries is of growing interest in relation to the evolution of asteroids under the influence of the YORP effect. Observations suggest that up to 10% of all NEAs are contact binaries, or bi-lobed objects, characterized by two distinct components.[74] The origin of such configurations is unknown. Although studies of the YORP effect suggest that a contact binary could easily separate into a stable binary system,[75] the mechanism for forming the original contact binary is unclear. It could be the result of YORP acting on a single progenitor to form a secondary that grows to be nearly the size of the primary. At that point tidal effects, and binary YORP effects (so-called "BYORP") could cause the secondary to spiral inward, gently impacting the primary and creating a contact binary shape.[76] The details of these processes are still under investigation.

**Conclusions**

Collisional processes have continuously shaped asteroids, and the recent joining of SPH numerical simulations with *N*-body gravitational simulations has vastly improved our understanding of how the population of small asteroids has evolved over time. Gravitational reaccumulation following the catastrophic disruption of Main Belt asteroids provides a good match to known asteroid families, and may be the source of small gravitational aggregates that migrate into the near-Earth population. The scaling laws for disruption, and actual laboratory experiments, are constantly being refined to account for different types of material, which correspond different populations of bodies. Upcoming all-sky surveys will dramatically increase the numbers of known asteroids, bringing greater definition to asteroid families and the size distribution in each population of the Solar System. Together, more complete models of the Solar System evolution should emerge.

The overall importance of rotational disruption for the evolution of small asteroids is still being pursued. It is now clear that a significant fraction of small asteroids, at least in the inner Main Belt and near-Earth populations, have satellites, and that most of these are probably formed by the YORP effect. Whether contact binaries, which account for nearly 10% of NEAs, have a rotational disruption origin is an open question. Mechanisms for the overall alteration of a body's surface and internal properties (spectral reflectance, thermal inertia, bulk porosity and density, spin poles etc.) will no doubt be further refined with improved models and observations in the near future.

**Acknowledgements:**
We would like to thank the various funding agencies who have supported the authors during the course of their work. KJW is currently supported by the Henri Poincaré fellowship at the Observatoire de la Côte d'Azur, Nice, France. PM acknowledges support of the French Programme National de Planétologie and DCR is acknowledges support from NNX08AM39G (NASA) and AST0708110 (NSF).

**Biographies:**